\documentclass[12pt,twoside]{article}
\usepackage{fleqn,espcrc1}

\usepackage{graphicx}
\usepackage[figuresright]{rotating}

\newcommand{\AmS}{{\protect\the\textfont2
  A\kern-.1667em\lower.5ex\hbox{M}\kern-.125emS}}

\title{Thermal bremsstrahlung probing the thermodynamical state of multifragmenting systems}

\author{D.G. d'Enterria\address{ \small{GANIL, BP 5027, 14076 Caen Cedex 5, France}\\
        $^b$ \small{Kernfysisch Versneller Instituut, 9747 AA Groningen, The Netherlands}\\
        $^c$ \small{Institute of Nuclear Physics, 250 68 \v Re\v z, Czech Republic}\\
        $^d$ \small{IFIC, Universitat de Val\`encia-CSIC, Dr. Moliner 50, 46100 Burjassot, Spain}\\
        $^e$ \small{Oak Ridge National Laboratory, Oak Ridge, TN 37831, USA}\\
        $^f$ \small{II. Physikalisches Institut, Universit\"at Gie\ss en, 35392 Gie\ss en, Germany}\\
        $^g$ \small{Grup de F\'{\i}sica Radiacions, Universitat Aut\`onoma de Barcelona, 08193 Bellaterra,
        Catalonia}\\}$^{,g}$\thanks{Present address: SUBATECH, 4 rue A. Kastler, 44307 Nantes Cedex 3,
        France}, L. Aphecetche$^a$$^*$, A. Chbihi$^a$, H. Delagrange$^a$$^*$, J. D\'{\i}az$^d$,
        M.J. van Goethem$^b$, M. Hoefman$^b$,  H. Huisman$^b$, A. Kugler$^c$, H. Loehner$^b$,
        G. Mart\'{\i}nez$^a$$^*$, R. Ortega$^g$, R. Ostendorf$^b$, S. Schadmand$^f$,
        Y. Schutz$^a$$^*$, R. Siemssen$^b$, D. Stracener$^e$, P. Tlusty$^c$,
        R. Turrisi$^a$\thanks{Present address: Universit\`a di Padova-INFN,  via Marzolo 8, 35131 Padova, Italy},
        M. Volkerts$^b$, V. Wagner$^c$, H. Wilschut$^b$, and N. Yahlali$^d$.
        }

\begin{document}

\maketitle

\begin{abstract}
Inclusive and exclusive hard-photon (E$_\gamma\,>$ 30 MeV)
production in five different heavy-ion reactions
($^{36}$Ar+$^{197}$Au, $^{107}$Ag, $^{58}$Ni, $^{12}$C at 60{\it A}
MeV and $^{129}$Xe+$^{120}$Sn at 50{\it A} MeV) has been studied
coupling the TAPS photon spectrometer with several
charged-particle multidetectors covering more than 80\% of 4$\pi$.
The measured spectra, slope parameters and source velocities as
well as their target-dependence, confirm the existence of thermal
bremsstrahlung emission from secondary nucleon-nucleon collisions
that accounts for roughly 20\% of the total hard-photon yield. The
thermal slopes are a direct measure of the temperature of the
excited nuclear systems produced during the reaction.
\end{abstract}

\vspace{0.5cm}
Nucleus-nucleus collisions constitute the only mean to study in the
laboratory the thermodynamical properties of hot and dense nuclear
matter and, consequently, to determine the nuclear
equation-of-state \cite{enterria:Peil94}. Heavy-ion (HI) reactions
at bombarding energies between 20{\it A} MeV and 100{\it A} MeV
lead to the formation of chunks of nuclear matter at subnuclear
densities and moderate excitation energies in the vicinity of the
predicted transition from the Fermi liquid phase to the nucleon gas
phase \cite{enterria:Poch97}. HI collisions are however dynamical
processes involving finite and transient systems. This fact renders
difficult the extraction of the thermostatistical properties of
infinite nuclear matter at equilibrium. As a matter of fact,
several key questions still remain open: Is thermodynamical
equilibrium attained? If so, what is the temperature of the excited
nuclear systems produced? What is the time-scale of nuclear
break-up? Is multifragmentation a signal of the liquid-gas phase
transition and/or of the passage through the spinodal region of the
phase diagram?
\\
To answer these questions, precise experimental probes of the
phase-space evolution of the HI collision are needed. This is the
primary motivation for the investigation of ``elementary" particle
production (hard photons, dileptons or mesons)
\cite{enterria:Cass90}. Despite their very small production
cross-sections, such energetic particles convey valuable
information about the stages of the reaction in which they are
created. Among them, photons are very clean probes since, due to
their weak electromagnetic coupling to nucleons, they have a small
probability of interacting with the surrounding medium and provide
a faithful image of the emission source. Hard-photons with
$E_\gamma\,>$ 30 MeV have been consistently interpreted as issuing
from the bremsstrahlung scattering of protons against neutrons,
$pn\,\rightarrow\,pn\gamma$, within the first 50 fm/c of the
reaction \cite{enterria:Cass90,enterria:Nife90}. Such prompt
photons provide thus valuable information about the {\it two-body}
dissipation mechanism in the compressed and pre-equilibrium phase
of the reaction. During the last 5 years, however, it has been
experimentally proven that the production of hard-photons
exclusively through first-chance $NN$ collisions needed to be
reconsidered as the existence of a bremsstrahlung emission
component of thermal origin emerged
\cite{enterria:Mart95,enterria:Marq95,enterria:Schu97}. To confirm
the existence of this second-chance bremsstrahlung emission and to
exploit its characteristics to extract the thermodynamical
properties (temperature, density) of the hot nuclear source(s), two
campaigns of the TAPS collaboration were carried out in 1997 and
1998 at the KVI and GANIL facilities. These experiments coupled the
TAPS photon spectrometer with several charged-particle
multidetectors for $\gamma$-particle coincident detection. The
heavy-ion reactions studied were $^{36}$Ar+$^{197}$Au, $^{107}$Ag,
$^{58}$Ni, $^{12}$C at 60{\it A} MeV and $^{129}$Xe+$^{120}$Sn at
50{\it A} MeV.

\section{EXPERIMENTAL RESULTS}

\begin{figure}[htbp]
\includegraphics[width=8cm]{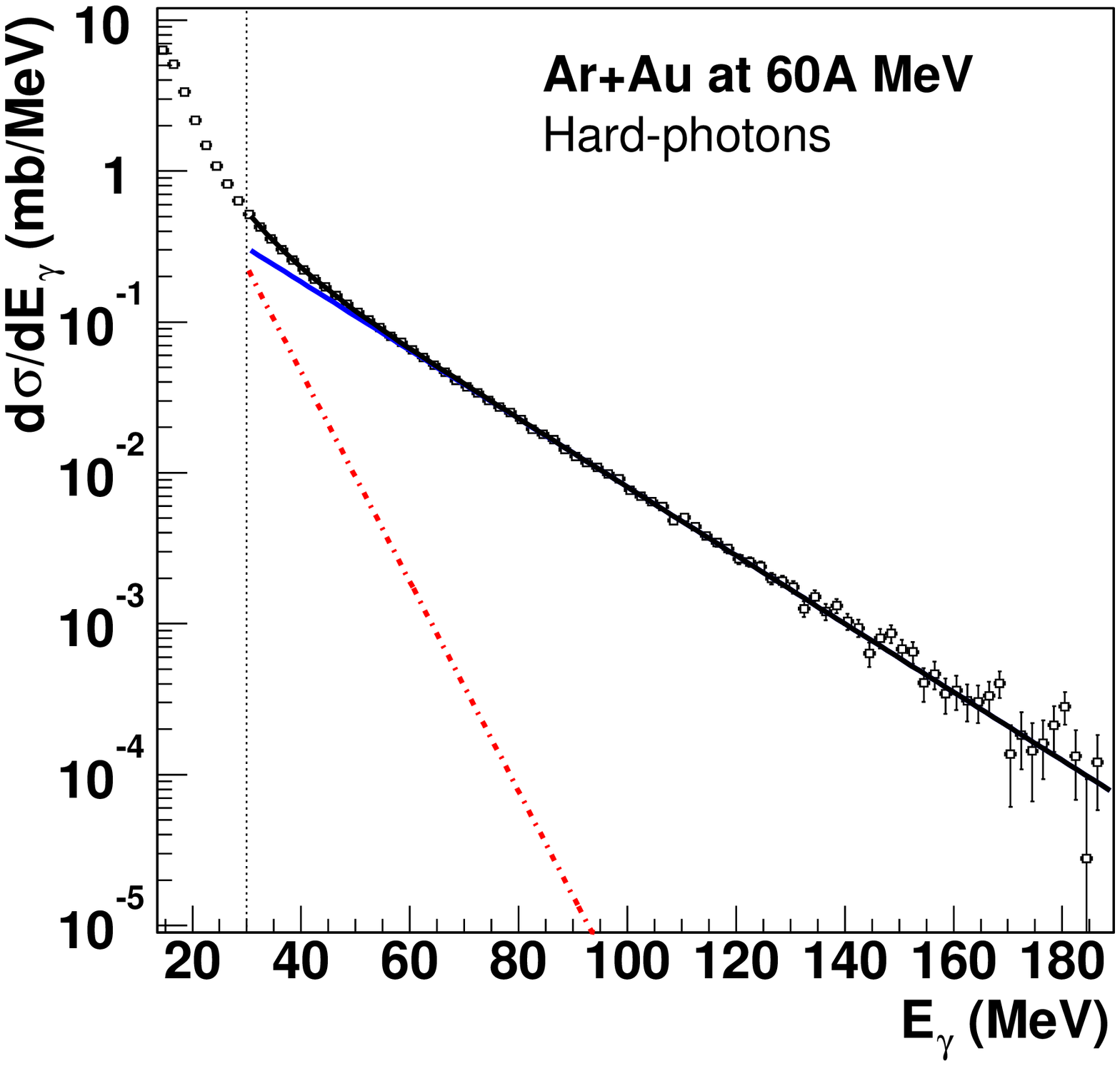}\hfill
\includegraphics[width=8cm]{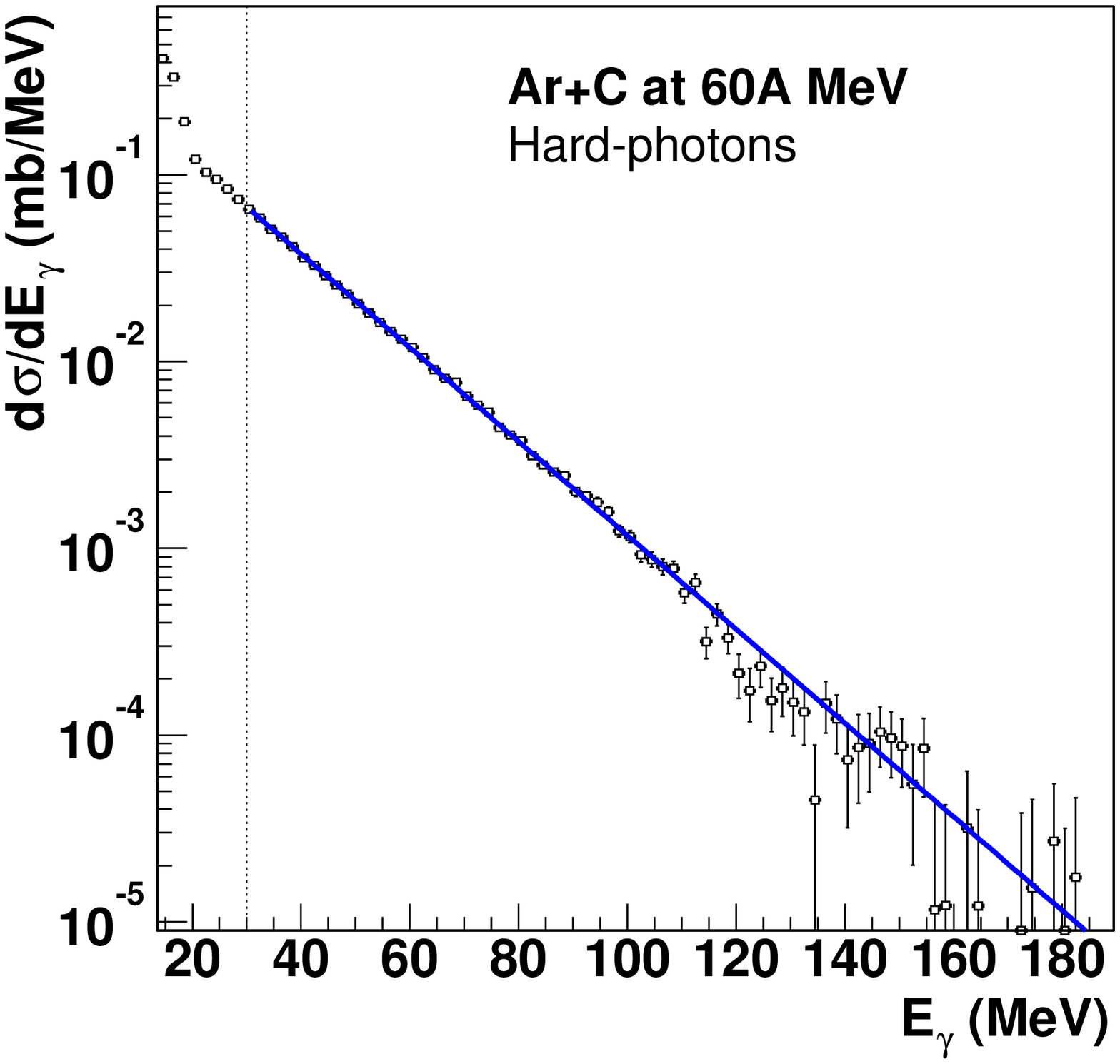}
 \caption{Hard-photon (E$_\gamma\,>$ 30 MeV) spectra measured for the
heaviest ($^{36}$Ar+$^{197}$Au, {\it left}) and lightest
($^{36}$Ar+$^{12}$C, {\it right}) systems and fitted, according to
Eq. (\ref{eq:enterria:1}), to the sum of two exponential
distributions: a {\it direct} (solid line) and a {\it thermal} one (dashed
line).}
 \label{fig:enterria:1}
\end{figure}

The inclusive photon energy spectra in the $NN$ CM frame have been
obtained after correction for the detector response function and
subtraction of the cosmic and radiative $\pi^0$-decay contributions
(Fig. \ref{fig:enterria:1}). The spectrum of the
$^{36}$Ar+$^{197}$Au system\footnote{As well as that of the
$^{36}$Ar+$^{107}$Ag, $^{58}$Ni and $^{129}$Xe+$^{120}$Sn
reactions, not shown in Fig. \ref{fig:enterria:1}.} features two
distinct exponential distributions with different slopes, and can
be described by \cite{enterria:Mart95}:
\begin{equation}
\label{eq:enterria:1}
\frac{d\sigma}{d E_\gamma}=K_d\:e^{-E_\gamma/E_0^d}+K_t\:e^{-E_\gamma/E_0^t}
\end{equation}

For the four heavier targets the slope parameters of the ``direct"
component ($E_0^d\,\approx$ 15 - 20 MeV) are two to three times
larger than the ``thermal" ones ($E_0^t\,\approx$ 6 - 9 MeV) and
the contribution of thermal hard-photons represents a 15\% - 25\%
of the total hard-photon yield \cite{enterria:Ente00,enterria:Orte00}. 
No thermal component is apparent in the photon spectrum of the
small $^{36}$Ar+$^{12}$C projectile-target combination and pure
direct bremsstrahlung clearly accounts for the whole photon
emission above $E_\gamma$ = 20 MeV (Fig. \ref{fig:enterria:1},
\textit{right}).

\begin{figure}[htbp]
\begin{center}
\includegraphics[width=9.0cm]{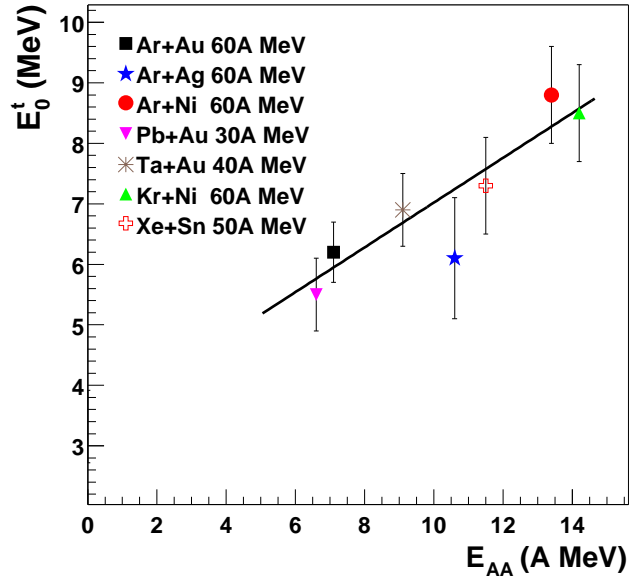}
 \end{center}
 \caption{Thermal hard-photon slopes as a function of the
 (Coulomb-corrected) nucleus-nucleus center-of-mass energy for the
 reactions studied at KVI in 1997 \cite{enterria:Ente00}, and at
 GANIL in 1998 \cite{enterria:Orte00} and 1992
 \cite{enterria:Schu97}. The solid line is a linear fit to the
 data.}
  \label{fig:enterria:2}
\end{figure}

The slopes of the direct component, $E_0^d$, follow the known
linear dependence with the {\it projectile energy per nucleon in
the laboratory} \cite{enterria:Ente00} as expected for
pre-equilibrium emission in prompt $NN\gamma$ collisions
\cite{enterria:Nife90} (the large values of the slope reflect the
coupling of the beam energy with the intrinsic Fermi momentum of
the colliding nucleons). The thermal hard-photon slopes, $E_0^t$,
at variance, scale with the {\it total energy in the
nucleus-nucleus center-of-mass} (Fig. \ref{fig:enterria:2}). This
property points to a thermal process taking place during later
stages of the reaction after dissipation of the incident kinetic
energy among internal degrees of freedom over the whole system in
the $AA$ center-of-mass (the lower slope values reflecting the less
energy available in secondary $NN\gamma$ collisions).

The interpretation of the second component of the hard-photon
spectrum as being emitted during later stages of the reaction in a
thermal process is also confirmed by the study of the
(Doppler-shifted) laboratory angular distributions
\cite{enterria:Ente00}. The hard-photon angular distributions can
be well interpreted assuming an emission from a source with slope
parameter $E_0^d$ moving with $\beta_S^d\,\approx\,\beta_{NN}$ plus
an isotropic source with slope parameter $E_0^t$ and
$\beta_S^t\,\approx\,\beta_{AA}$, with the ratios of thermal to
direct intensities being fixed by the energy spectra. This result
is consistent with {\it direct} hard-photons being emitted from the
$NN$ center-of-mass, and {\it thermal} hard-photons being emitted
isotropically from a slowed-down source moving with the $AA$
center-of-mass velocity.

\section{THERMODYNAMICAL PROPERTIES OF THE NUCLEAR SOURCES}

The existence of a thermal mechanism accounting for part of the
total hard-photon yield, justifies the use of a thermal
bremsstrahlung model \cite{enterria:Neuh87} to extract the
thermodynamical properties of the radiating nuclear systems. Such a
model predicts thermal photon spectra basically exponential in the
region $E_\gamma$ = 30 - 80 MeV in agreement with our data. The
slopes $E_0^t$ of such exponential spectra are linearly correlated with the
local temperature $T$ of the nuclear source according to:
\begin{equation}
\label{eq:T vs E0t}
T(\mbox{MeV})\,=\,a\cdot E_0^t(\mbox{MeV})\, -\, b\,
\quad \mbox{ with $a$ = 0.75 $\pm$ 0.05  and $b$ = -0.65 $\pm$ 0.05 MeV}
\end{equation}

The correlation of the temperatures obtained using eq. (\ref{eq:T vs
E0t}), in the range $T$ = 4 - 6 MeV, with the excitation energies
attained in each reaction ($\mbox{{\large $\epsilon$}}^\star$ =
4{\it A} - 10{\it A} MeV) yields a ``caloric curve" which shows a
slightly increasing ``plateau" \cite{enterria:Ente00b}. Such a
trend, observed by the ALADIN collaboration, was interpreted as a
signal of the nuclear liquid-gas phase transition
\cite{enterria:Poch97}. This observation disagrees with the higher
apparent temperatures obtained using the slopes of the
(Maxwell-Boltzmann) kinetic energy distributions of different
light-particles ($p,\;n$ or $\alpha$) \cite{enterria:Ma97}.
However, at variance with ALADIN data, we measure temperatures for
nuclear systems around the saturation density ($NN\gamma$
collisions only take place with sizeable cross-sections around or
above $\rho_0$). We conclude that the hot radiating nuclear
residues are more likely in a liquid-gas coexistence phase in an
``evaporation"-like scenario, than undergoing a simultaneous
breakup in a dilute state.


\begin{thebibliography}{1234}
\bibitem{enterria:Peil94}G. Peilert, H. St\"{o}cker and W. Greiner,  Rep. Prog. Phys. {\bf 57} (1994) 533.
\bibitem{enterria:Poch97}J. Pochodzalla,  Prog. Part. Nucl. Phys. {\bf 39} (1997) 443.
\bibitem{enterria:Cass90}W.~Cassing, V.~Metag, U.~Mosel and K.~Niita,  Phys. Rep. {\bf 188} (1990) 363.
\bibitem{enterria:Nife90}H.~Nifenecker and J.~Pinston,  Ann. Rev. Nucl. Part. Sci. {\bf 40} (1990) 113.
\bibitem{enterria:Mart95}G.~Mart\'{\i}nez {\it et al.},  Phys. Lett. {\bf B349} (1995) 23.
\bibitem{enterria:Marq95}F.M.~Marqu\'es {\it et al.},  Phys. Lett. {\bf B349} (1995) 30.
\bibitem{enterria:Schu97}Y.~Schutz {\it et al.},  Nucl. Phys. {\bf A622} (1997) 405.
\bibitem{enterria:Ente00}D.G.~d'Enterria,  Proc. V TAPS Workshop, \v Re\v z, Sept. 1999, nucl-ex/0007005.
\bibitem{enterria:Orte00}R. Ortega,  Proc. V TAPS Workshop, \v Re\v z, Sept. 1999, Czech Jour. Phys.
\bibitem{enterria:Neuh87}D.~Neuhauser and S.E.~Koonin,  Nucl. Phys. {\bf A462} (1987) 163.
\bibitem{enterria:Ente00b}D.G.~d'Enterria, PhD thesis, U.A. Barcelona and U. Caen, Mars 2000.
\bibitem{enterria:Ma97}Y.-G.~Ma {\it et al.}, Phys. Lett. {\bf B390} (1997) 41.
\end{thebibliography}
\end{document}